\documentclass{SciPost}

\hypersetup{
    colorlinks,
    linkcolor={red!50!black},
    citecolor={blue!50!black},
    urlcolor={blue!80!black}
}

\usepackage[bitstream-charter]{mathdesign}
\DeclareSymbolFont{usualmathcal}{OMS}{cmsy}{m}{n}
\DeclareSymbolFontAlphabet{\mathcal}{usualmathcal}

\fancypagestyle{SPstyle}{
\fancyhf{}
\lhead{\colorbox{scipostblue}{\bf \color{white} ~SciPost Physics Proceedings }}
\rhead{{\bf \color{scipostdeepblue} ~Submission }}

\fancyfoot[C]{\textbf{\thepage}}
}

\begin{document}

\pagestyle{SPstyle}

\begin{center}{\Large \textbf{\color{scipostdeepblue}{
%%%%%%%%%% TODO: Write your article's title here
%Article Title, as descriptive as possible, ideally fitting in two lines (approximately 150 characters) or less\\
On electroweak metastability and Higgs inflation
%%%%%%%%%% END TODO: TITLE
}}}\end{center}

\begin{center}\textbf{
%%%%%%%%%% TODO: AUTHORS
% Write the author list here. 
% Use (full) first name (+ middle name initials) + surname format.
% Separate subsequent authors by a comma, omit comma and use "and" for the last author.
% Mark the corresponding author(s) with a superscript symbol in this order
% \star, \dagger, \ddagger, \circ, \S, \P, \parallel, ...
Isabella Masina\textsuperscript{1$\star$},
Mariano Quiros \textsuperscript{2$\dagger$} 
%%%%%%%%%% END TODO: AUTHORS
}\end{center}

\begin{center}
%%%%%%%%%% TODO: AFFILIATIONS
% Write all affiliations here.
% Format: institute, city, country
{\bf 1} Ferrara University and INFN Sez.\,Ferrara, Dept.\,of Physics and Earth Science, Via Saragat 1, 44122 Ferrara, Italy
\\
{\bf 2} Institut de F\'{\i}sica d'Altes Energies (IFAE) and The Barcelona Institute of  Science and Technology (BIST), Campus UAB, 08193 Bellaterra (Barcelona) Spain
%%%%%%%%%% END TODO: AFFILIATIONS
%%%%%%%%%% TODO: EMAIL
% Provide email address of corresponding author(s)
\\[\baselineskip]
$\star$ \href{mailto:email1}{\small masina@fe.infn.it}\,,\quad
$\dagger$ \href{mailto:email2}{\small quiros@ifae.es}
%%%%%%%%%% END TODO: EMAIL
\end{center}

\definecolor{palegray}{gray}{0.95}
\begin{center}
\colorbox{palegray}{
  \begin{tabular}{rr}
  \begin{minipage}{0.36\textwidth}
    \includegraphics[width=60mm,height=1.5cm]{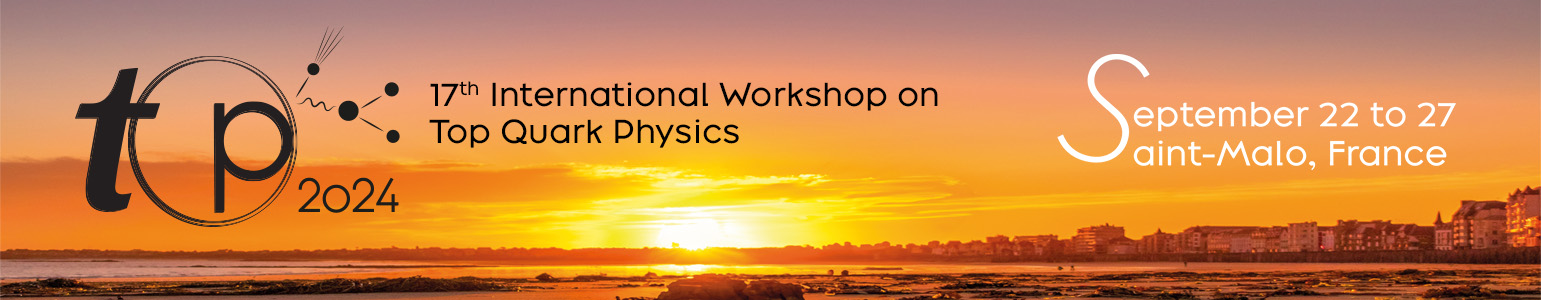}
  \end{minipage}
  &
  \begin{minipage}{0.55\textwidth}
    \begin{center} \hspace{5pt}
    {\it The 17th International Workshop on\\ Top Quark Physics (TOP2024)} \\
    {\it Saint-Malo, France, 22-27 September 2024
    }
    \doi{10.21468/SciPostPhysProc.?}\\
    \end{center}
  \end{minipage}
\end{tabular}
}
\end{center}

\section*{\color{scipostdeepblue}{Abstract}}
\textbf{\boldmath{%
%%%%%%%%%% TODO: ABSTRACT
%The abstract is in boldface, and should fit in 8 lines. It should be written in a clear and accessible style, emphasizing the context, the problem(s) studied, the methods used, the results obtained, the conclusions reached, and the outlook. You can add a table contents, recommended if your paper is more than 6 pages long.
For the central values of the relevant experimental inputs, that is the strong coupling constant and the top quark and Higgs masses,
the effective Higgs potential displays two minima, one at the electroweak scale and a deeper one at high energies.
We review the phenomenology of the Higgs inflation model, extending the Standard Model to include a non-minimal coupling to gravity;
as recently shown \cite{Masina:2024ybn}, even configurations that would be metastable in the Standard Model, 
become viable for inflation if the non-minimal coupling is large enough to flatten the Higgs potential at field values below the barrier between the minima. 
%%%%%%%%%% END TODO: ABSTRACT
}}

\vspace{\baselineskip}

%%%%%%%%%% BLOCK: Copyright information
% This block will be filled during the proof stage, and finilized just before publication.
% It exists here only as a placeholder, and should not be modified by authors.
\noindent\textcolor{white!90!black}{%
\fbox{\parbox{0.975\linewidth}{%
\textcolor{white!40!black}{\begin{tabular}{lr}%
  \begin{minipage}{0.6\textwidth}%
    {\small Copyright attribution to authors. \newline
    This work is a submission to SciPost Phys. Proc. \newline
    License information to appear upon publication. \newline
    Publication information to appear upon publication.}
  \end{minipage} & \begin{minipage}{0.4\textwidth}
    {\small Received Date \newline Accepted Date \newline Published Date}%
  \end{minipage}
\end{tabular}}
}}
}
%%%%%%%%%% BLOCK: Copyright information

%%%%%%%%%% TODO: LINENO
% For convenience during refereeing we turn on line numbers:
%\linenumbers
% You should run LaTeX twice in order for the line numbers to appear.
%%%%%%%%%% END TODO: LINENO

%%%%%%%%%% TODO: TOC 
% Guideline: if your paper is longer that 6 pages, include a TOC
% To remove the TOC, simply cut the following block
\vspace{10pt}
\noindent\rule{\textwidth}{1pt}
\tableofcontents
\noindent\rule{\textwidth}{1pt}
\vspace{10pt}
%%%%%%%%%% END TODO: TOC

%%%%%%%%% TODO: CONTENTS 
% Write your article contents here, starting from first \section.
% An example structure is given below.

\section{Introduction}
\label{sec:intro}
% TODO: write your article here.
%The stage is yours. Write your article here.
%The bulk of the paper should be clearly divided into sections with short descriptive titles, including an introduction and a conclusion.

In the Standard Model (SM), the central values of the relevant experimental inputs -- the strong coupling constant with five flavors, 
$\alpha_s^{(5)}$, the top quark pole mass, $m_t$, and the Higgs mass, $m_H$ -- suggest that the electroweak vacuum is likely to be metastable rather than stable.
The shape of the Higgs effective potential at high energy is relevant in view of the possible role of the Higgs field as the inflaton;
for this sake, a region where the Higgs potential becomes sufficiently flat, for large enough values of the Higgs field, to meet the slow-roll conditions would be required.

By introducing a non-minimal coupling to gravity, $\xi$, that flattens
%\footnote{For an alternative approach postulating flatness, but without the non-minimal coupling, see\,\cite{Hamada:2013mya}.} 
the Higgs potential at field values larger than about $M_P/\sqrt{\xi}$,
where $M_P$ is the reduced Planck scale, the Higgs might successfully play the role of the inflaton~\cite{BezHiggs}.
In this contribution, based on \cite{Masina:2024ybn}, we review the phenomenology of Higgs inflation, including would-be metastable configurations.
% $M_P=1/\sqrt{8\pi G_N}$

%%%%%%%%%%%
\section{Metastability of the Standard Model Higgs potential}

Consider the SM Higgs doublet, $H=(0, \phi+v)^T/\sqrt{2}$, where $v$ is the electroweak vev. Via the Renormalization Group Equations, it is possible to extrapolate the SM Higgs effective potential, $V_{\rm eff}$, at high energies. The most relevant experimental inputs to determine the shape of the Higgs potential are $m_t$, $m_H$ and $\alpha_s^{(5)}$. 
According to their values, three scenarios can be found: instability, metastability and stability.

The program of discriminating these scenarios started in the fall of the '70s ~\cite{Hung:1979dn, Cabibbo:1979ay, Casas:1994qy, Froggatt:1995rt}, after the prediction of the top quark in 1973, but much before its discovery in 1995. The 2012 discovery of the Higgs boson by LHC %, with mass \textcolor{red}{$m_H = 125.2\pm 0.11$ GeV}, 
triggered theoretical improvements, and the extrapolation reached the NNLO accuracy: 2-loop in the effective potential and matching conditions, and 3-loop in running. 
Still, the ambiguity was left among stability and metastability, the latter being slightly favored (see {\it e.g.}\cite{Degrassi, Masina:2012tz, Buttazzo:2013uya, Bednyakov:2015sca, Hiller:2024zjp}); the difference between instability and metastability requires an analysis of the tunneling probability from the false to the true minima. 
Even in the future, it will be difficult to exclude stability~\cite{Franceschini:2022veh}. 
\begin{figure}[htb]
	\centering
	\includegraphics[width=0.48 \textwidth]{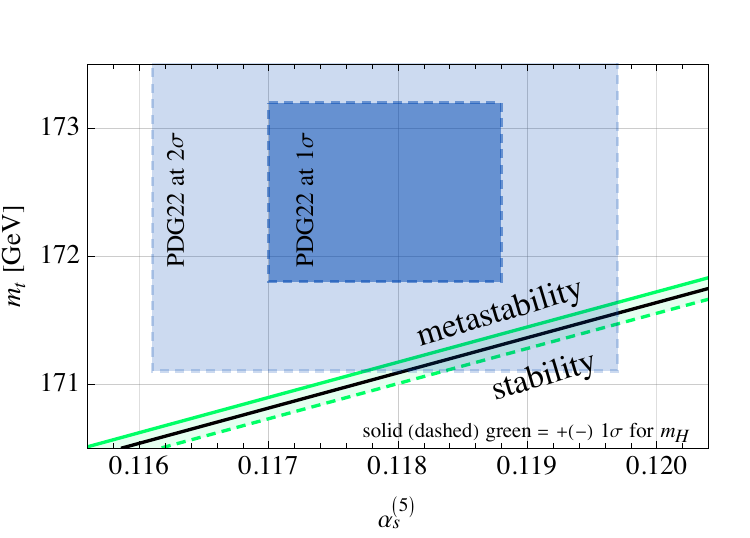}  
	\includegraphics[width=0.5 \textwidth]{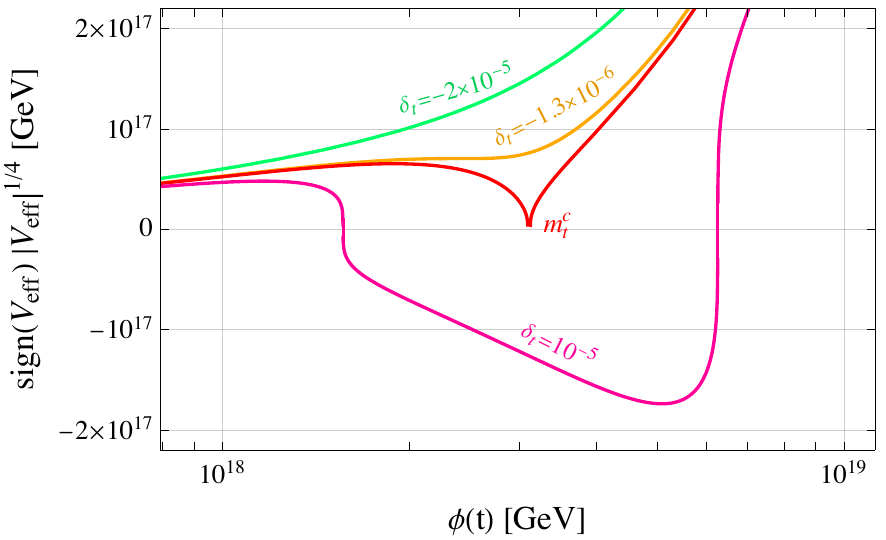} 
	\caption{ 
	Left: the black line separates the regions of stability and metastability; 
	the green solid and dashed lines display the effect of a positive and negative $1\,\sigma$ variation in $m_H$; 
	the shaded blue rectangles display the $1\,\sigma$ and $2\,\sigma$ allowed ranges for $\alpha_s^{(5)}$ and $m_t$ ~\cite{PDG2022}. 
        Right: Higgs effective potential configurations near criticality. From~\cite{Masina:2024ybn}.}
\label{fig-meta}
\end{figure}

%Bednyakov:2015sca

Taking central values for $\alpha_s^{(5)}$ and $m_H$~\cite{PDG2022}, the (critical) value of the top mass for which the NNLO Higgs effective potential displays two degenerate minima is $m_t^c \approx 171.0588$ GeV, as discussed {\it e.g.}~in Ref.~\cite{Masina:2024ybn}. 
The left plot in Fig.\,\ref{fig-meta} shows the regions where the electroweak vacuum is stable or metastable.
%; the green solid and dashed lines display the effect of a positive and negative $1\,\sigma$ variation in $m_H$, 
%while the shaded blue rectangles display the $1\,\sigma$ and $2\,\sigma$ allowed ranges for $\alpha_s^{(5)}$ and $m_t$~\cite{PDG2022}. 
  
It is also interesting to study the shape of the Higgs potential around criticality, as shown in the right plot of Fig.\,\ref{fig-meta}.
The critical configuration is shown in red, and one can see that the high energy vacuum is located at field values $\phi(t)$ which are close to the Planck energy scale.  
It is useful to define 
\begin{equation}
\delta_t= m_t/m_t^c -1,
\end{equation} 
and explore possible shapes close to the critical one, as for instance the inflection point configuration (in orange), achieved with 
$\delta_t = -1.3 \times 10^{-6}$, 
and the configurations corresponding to $\delta_t=-2\times 10^{-5}$ and $\delta_t=10^{-5}$.

%%%%%%%%%%%
\section{Inflation, new physics, and the Higgs field}

The basic idea of inflation is to introduce a homogeneous scalar field, called inflaton.
If, for some reason, there has been a period in which the Hubble rate $H(t)$ was dominated by a positive nearly constant potential $V(\phi)$,
then 
\begin{equation}
\left(\frac{\dot a(t)}{a(t)}\right)^2 = H^2(t) \approx \frac{8 \pi V(\phi)}{3 M_P^2}, 
\end{equation}
where $a(t)$ is the scale factor;
a period of exponential expansion is thus achieved, $a(t) \propto e^{H t}$. The latter might address issues like the Universe flatness, isotropy and homogeneity, etc.
If the potential $V(\phi)$ is sufficiently flat, the slow-roll conditions are met, 
\begin{equation}
\epsilon\equiv \frac{M_P^2}{2}\left(\frac{V'}{V} \right)^2\ll 1 \,, \quad |\eta|\equiv M_P^2\frac{|V''|}{V}\ll 1, 
\end{equation}
and the cosmological observables are:
\begin{equation}
\Delta_R^2\simeq V/(24\pi^2 M_P^4\epsilon)\simeq 2.1\times 10^{-9}, \quad 
n_s\simeq 1+2\eta-4\epsilon, \quad
r=16\epsilon,
\label{eq-cosobs}
\end{equation}
where $\Delta_R^2$ is the amplitude of density perturbations, $n_s$ the inflaton spectral index, and $r$ the tensor-to-scalar ratio.
The inflationary period should end after about $N=60$ e-folds, leading to matter production via the so-called reheating process.

As the Higgs is the only elementary scalar found, could it be involved in primordial inflation?
Let us start from top to bottom in the right plot of Fig.~\ref{fig-meta}, that is from stable to metastable configurations: 
i) If the Higgs potential is ever increasing, it is also too steep and slow-roll cannot take place;
ii) In the case of an inflection point, a bit of slow roll is found, but not for enough e-folds \cite{Isidori:2007vm};
iii) Critical and metastable configurations clearly do not work.

\begin{figure}[h!]
	\centering
	 \includegraphics[width=0.7 \textwidth]{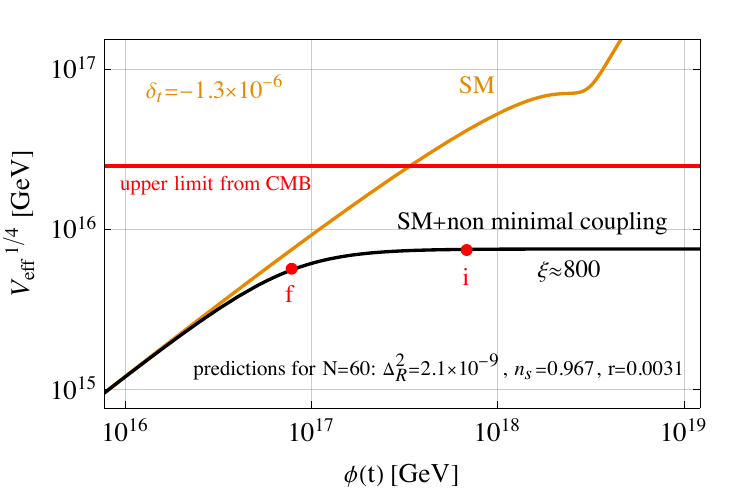}   
		\caption{The upper curve is the SM inflection point configuration. The lower (black) curve shows how the same configuration gets modified by including $\xi \approx 800$; the cosmological predictions agree with CMB data~\cite{PDG2022}, and the red dots signal the beginning (i) and end (f) of inflation. Adapted from~\cite{Masina:2024ybn}.}
	\label{fig-Hinfl}
\end{figure}

As a variation of i), one might envisage the possibility that the Higgs is trapped in a shallow vacuum (old inflation type), and  another field acts as inflaton.
This scenario cannot account for the observed cosmological parameters~\cite{Iacobellis:2016eof, Masina:2018ejw}: as shown in Fig.~\ref{fig-Hinfl}, the value of the Higgs effective potential at the stationary point exceeds the upper limit $V_{\rm eff}^{1/4} \lesssim 2.5 \times 10^{16}$ GeV~\cite{Masina:2024ybn}, derived from CMB data by combining,
via Eq.~(\ref{eq-cosobs}), the observed value of $\Delta_R^2$ with the upper bound on $r \lesssim 0.036$~\cite{PDG2022}.

The general lesson is that new physics beyond the SM is needed.
For instance: something that flattens the Higgs potential at some $\phi$, so that the Higgs itself is the inflaton.
Bezrukov and Shaposhnikov~\cite{BezHiggs} showed that is precisely what happens to $V_{\rm eff}$
by adding to the SM Lagrangian, $\mathcal{L}_{\rm SM}$, a non-minimal gravitational coupling $\xi$ between the SM Higgs doublet $H$ and the Ricci scalar $R$.
The classical action for such Higgs inflation model is
\begin{equation}
\mathcal{S}=\int d^{4}x \, \sqrt{-g}\left[\mathcal{L}_{\rm SM}-\frac{M_P^{2}}{2} R -\xi\left| H\right|^{2} R\right]\,,
\label{eq-act}
\end{equation}
where $g$ is the determinant of the FLRW %Friedmann-Lem\^aitre-Robertson-Walker 
metric.
The effect of the introduction of the non-minimal coupling to gravity, $\xi$, is to flatten the Higgs potential at field values larger than about $M_P/\sqrt{\xi}$, so that slow-roll conditions are met, and the Higgs successfully plays the role of the inflaton. In the metric formulation and for $N=60$ e-folds, the cosmological predictions are:
$n_s-1\approx -2/N = -0.033$, $r\approx 12/N^2 = 0.0033$. Substituting the latter predictions in (\ref{eq-cosobs}), one gets a numerical value for the Higgs potential at the beginning of inflation, $V_i^{1/4} \approx 7.6 \times 10^{15}$ GeV; the value of $\xi$ is determined from the requirement of matching such a value.

The previous literature focused on ever increasing stable configurations. However, thanks to the flattening mechanism, even configurations that in the SM would display two minima -- like a shallow vacuum or slightly metastable configurations -- become viable for inflation~\cite{Masina:2024ybn}. 
%This happens for both the metric and the Palatini formalisms of gravity. %~\cite{Palatini:1919ffw}.
For instance, as shown in Fig.\,\ref{fig-Hinfl} for the metric formalism, for the same input parameters that would lead to an inflection point configuration in the SM, the inclusion of a non-minimal coupling, $\xi \approx 800$, suitably flattens the Higgs effective potential, leading to viable cosmological predictions for $\Delta_R^2$, $n_s$ and $r$~\cite{PDG2022}.

%%%%%%%%%%%
\section{Conclusions and perspectives}

\begin{figure}[h]
	\centering
  \includegraphics[width=0.9 \textwidth]{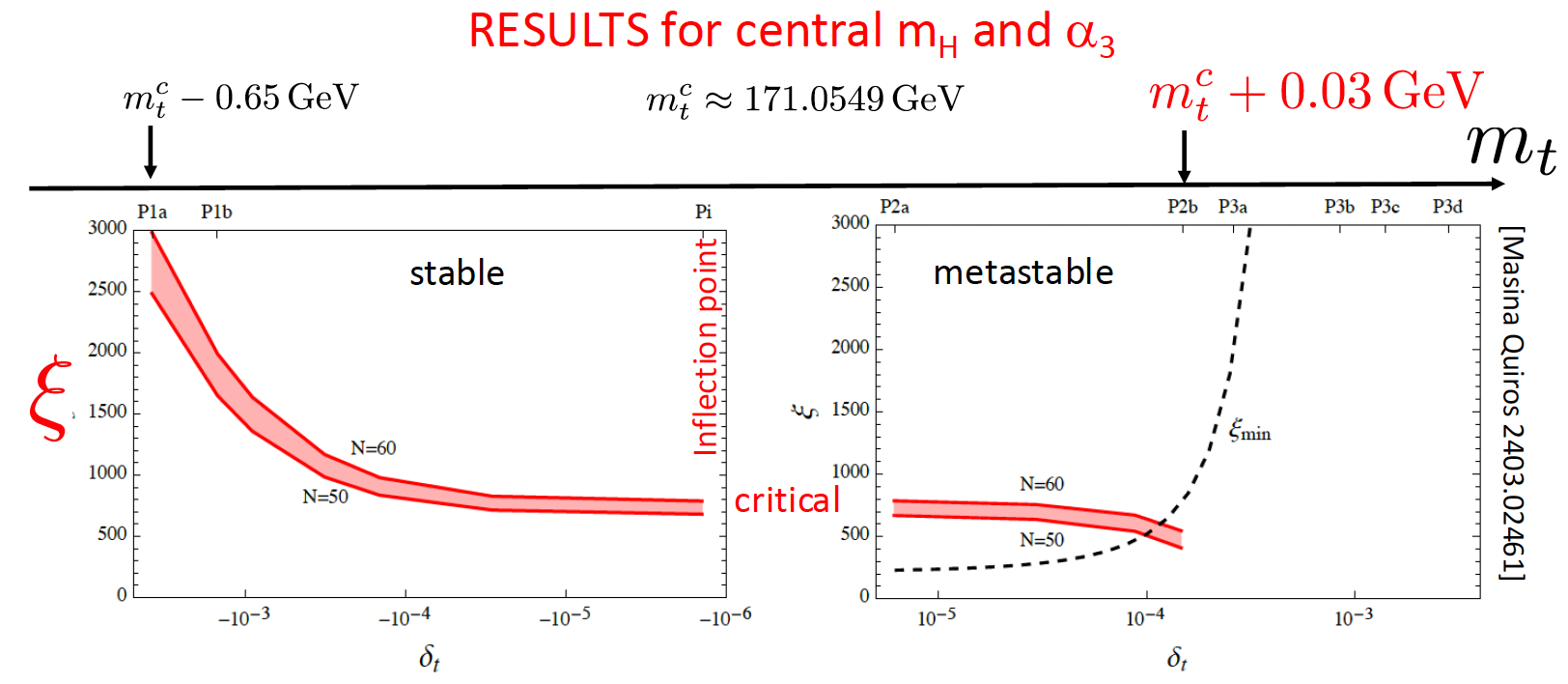}  
		\caption{The values of $\xi$ providing successful Higgs inflation. Adapted from~\cite{Masina:2024ybn}.}
	\label{fig-res}
\end{figure}

Fig.\,\ref{fig-res} summarizes the results of our analysis, by showing the values of $\xi$ providing successful Higgs inflation as a function of $\delta_t$, hence of the top quark mass. Notice that $\xi$ decreases as $m_t$ increases, and that for configurations close to metastability $\xi$ can be as small as about 500.
 
There are no roses without a thorn. In this context, the debated issues of unitarity and which formalism to adopt attracted much interest;
for a dedicated discussion about those aspects, we refer to~\cite{Masina:2024ybn}. Here we just mention that with $\xi$ as small as $500$, 
unitarity issues should not apply, and the inflationary dynamics is expected to be reliable. 

To conclude, a future better experimental determination of the top quark mass value would be extremely relevant and helpful for Higgs inflation models.

%%%

%In the list of references, cited papers \cite{1931_Bethe_ZP_71} should include authors, title, journal reference (journal name, volume number (in bold), start page) and most importantly a DOI link. For a preprint \cite{arXiv:1108.2700}, please include authors, title (please ensure proper capitalization) and arXiv link. If you use BiBTeX with our style file, the right format will be automatically implemented.
%
%All equations and references should be hyperlinked to ensure ease of navigation. This also holds for [sub]sections: readers should be able to easily jump to Section \ref{sec:another}.

%%%%%%%%
%\section{Another Section}
%\label{sec:another}
%
%
%\subsection{A note about figures}
%Figures should only occupy the stricly necessary space, in any case individually fitting on a single page. Each figure item should be appropriately labeled and accompanied by a descriptive caption. {\bf SciPost does not accept creative or promotional figures or artist's impressions}; on the other hand, technical drawings and scientifically accurate representations are encouraged.

%%%%%%%
%\section*{Acknowledgements}
%Acknowledgements should follow immediately after the conclusion.

%% TODO: include author contributions
%\paragraph{Author contributions}
%This is optional. If desired, contributions should be succinctly described in a single short paragraph, using author initials.

% TODO: include funding information
\paragraph{Funding information}
%Authors are required to provide funding information, including relevant agencies and grant numbers with linked author's initials. Correctly-provided data will be linked to funders listed in the \href{https://www.crossref.org/services/funder-registry/}{\sf Fundref registry}.
IM acknowledges partial support by the research project TAsP (Theoretical Astroparticle Physics) 
funded by the Istituto Nazionale di Fisica Nucleare (INFN). MQ is supported by grant PID2023-146686NB-C31 funded by MICIU/AEI/10.13039/501100011033/ and by FEDER, EU. 
IFAE is partially funded by the CERCA program of the Generalitat de Catalunya.

\begin{appendix}
\numberwithin{equation}{section}

%\section{About references}
%Your references should start with the comma-separated author list (initials + last name), the publication title in italics, the journal reference with volume in bold, start page number, publication year in parenthesis, completed by the DOI link (linking must be implemented before publication). If using BiBTeX, please use the style files provided  on \url{https://scipost.org/submissions/author_guidelines}. If you are using our LaTeX template, simply add
%\begin{verbatim}
%\bibliography{your_bibtex_file}
%\end{verbatim}
%at the end of your document. If you are not using our LaTeX template, please still use our bibstyle as
%\begin{verbatim}
%\bibliographystyle{SciPost_bibstyle}
%\end{verbatim}
%in order to simplify the production of your paper.

\end{appendix}

%%%%%%%%% END TODO: CONTENTS

%%%%%%%%%% TODO: BIBLIOGRAPHY
% Provide your bibliography here. You have two options:

%%% FIRST OPTION
% Write your entries here directly, following the example below, including:
% Author(s), Title, Journal Ref. with year in parentheses at the end, followed by the DOI number.

% \begin{thebibliography}{99}
%  \bibitem{1931_Bethe_ZP_71}
% H. A. Bethe, \textit{Zur Theorie der Metalle. i. Eigenwerte und Eigenfunktionen der linearen Atomkette}, Zeit. f{\"u}r Phys. \textbf{71}, 205 (1931), \doi{10.1007\%2FBF01341708}.
% \bibitem{arXiv:1108.2700}
% P. Ginsparg, \textit{It was twenty years ago today...}, (arXiv preprint) \doi{10.48550/arXiv.1108.2700}.
% \end{thebibliography}

%%% SECOND OPTION
% Use your bibtex library, formatted by the SciPost style file.
\bibliography{SciPost-Masina-BiBTeX.bib}

%%%%%%%%%% END TODO: BIBLIOGRAPHY

\end{document}